\documentclass[aps,preprint,prd,nofootinbib]{revtex4}
\usepackage{graphicx}
\usepackage{psfrag}

\begin{document}

\title{The Strong Decays of Orbitally Excited $B^{*}_{sJ}$ Mesons by
Improved Bethe-Salpeter Method}
\author{Zhi-Hui Wang$^{[1]}$, Guo-Li Wang$^{[1]}$\footnote{gl\_wang@hit.edu.cn}, Hui-Feng Fu$^{[1]}$, Yue Jiang$^{[1]}$\\}
\address{$^1$Department of Physics, Harbin Institute of
Technology, Harbin, 150001, China.}
 \baselineskip=20pt

\begin{abstract}
\vspace{4mm}
 We calculate the masses and the strong decays of
orbitally excited states $B_{s0}$, $B'_{s1}$, $B_{s1}$ and
$B_{s2}$ by the improved Bethe-Salpeter method. The predicted
masses of $B_{s0}$ and $B'_{s1}$ are $M_{B_{s0}}=5.723\pm0.280~
{\rm GeV}$, $M_{B'_{s1}}=5.774\pm0.330~ {\rm GeV}$. We calculate
the isospin symmetry violating decay processes $B_{s0}\to B_s \pi$
and $B'_{s1}\to B_s^* \pi$ through $\pi^0-\eta$ mixing and get
small widths. Considering the uncertainties of the masses, for
$B_{s0}$ and $B'_{s1}$, we also calculate the OZI allowed decay
channels: $B_{s0}\to B\bar K$ and $B'_{s1}\to B^*\bar K$. For
$B_{s1}$ and $B_{s2}$, the OZI allowed decay channels $B_{s1}\to
B^{*}\bar K$, $B_{s2}\to B\bar K$ and $B_{s2}\to B^{*}\bar K$ are
studied. In all the decay channels, the reduction formula, PCAC
relation and low energy theorem are used to estimate the decay
widths. We also obtain the strong coupling constants
$G_{B_{s0}B_s\pi}$, $G_{B_{s0}B\bar K}$, $G_{B'_{s1}B_s^*\pi}$,
$F_{B'_{s1}B_s^*\pi}$, $G_{B'_{s1}B^*\bar K}$, $F_{B'_{s1}B^*\bar
K}$, $G_{B_{s1}B^{*}\bar K}$, $F_{B_{s1}B^{*}\bar K}$,
$G_{B_{s2}B\bar K}$ and $G_{B_{s2}B^{*}\bar K}$.


\end{abstract}

\maketitle

\section{Introduction}

The heavy-light mesons play an important role in hadron physics.
During the past several years, many heavy-light mesons were
observed in experiments. In the Particle Data Group (PDG) table
\cite{pdg}, there are four $P$-wave charm states
$D_{0}^*(2400)^0$, $D_{1}(2420)^0$, $D_{1}(2430)^0$,
$D_{2}^*(2460)^{0}$, and two $P$-wave bottom states $B_1(5721)^0$,
$B_2^*(5747)^0$. For the $P$-wave bottom-strange states, $B_{s1}$
and $B_{s2}^*$ are observed by the CDF Collaboration in 2008
\cite{cdf}. Later the D0 Collaboration also reported $B_{s2}^*$
\cite{d0}. Meanwhile D0 Collaboration indicated that $B_{s1}$ was
not observed with the available data set.

In the heavy quark effective theory(HQET) \cite{hqet}, for the
heavy-light meson system, the angular momentum of light quark
$j_l$ is a good quantum number when the heavy quark have $m_Q\to
\infty$ limit. They are $j^P_l={\frac{1}{2}}^-$ $H$ doublet ($0^-,
1^-$) with orbital angular momentum $L=0$; $j^P_l={\frac{1}{2}}^+$
$S$ doublet ($0^+, 1^+$) and $j^P_l={\frac{3}{2}}^+$ $T$ doublet
($1^+, 2^+$)  with orbital angular momentum $L=1$. The D0 and CDF
indicated that $B_{s1}(5830)$ and $B_{s2}^*(5840)$ correspond to
the states with $J^P=1^+$ and $J^P=2^+$ in $T$ doublet
\cite{cdf,d0}. While for the $B^*_{sJ}$ state with $S$ doublet
$J^P=(0^+, 1^+)$ do not have the experimental evidence. For the
charm states, $D_{0}^*(2400)^0$ and , $D_{1}(2430)^0$ are the
$\frac{1}{2}^+$ $S$ doublet ($0^+, 1^+$), $D_{1}(2420)^0$ and
$D_{2}^*(2460)^{0}$ belong to $\frac{3}{2}^+$ $T$ doublet ($1^+,
2^+$). $B_1(5721)^0$ and $B_2^*(5747)^0$ also belong to
$\frac{3}{2}^+$ $T$ doublet ($1^+, 2^+$).

The observations of these $P$-wave mesons inspire our interest in
their nature. There are many theoretical approaches are used to
study their properties
\cite{wangzg,L4,L5,qq,16,liu,supp,zhao,aff,hill}. In this work, we
focus on the productions of $P$-wave charm states, bottom states
and bottom-strange states in exclusive semileptonic and
nonleptonic $B_c$ decays.

Since the discovery of $D_{s0}(2317)$ \cite{2317}, the heavy-light
orbitally excited states stimulated continued interesting
attentions. There are some special characters of these excited
states, for example, the mass of $D_{s0}(2317)$ is much smaller
than the prediction of the relativistic quark model \cite{godfrey}
which has been very successful, and it has a narrow decay width.
Though it is believed to be the orbitally excited state of $D_s$
by most of the physicists now, there have been some arguments
about its nature. It can be a conventional $c\bar s$ state
\cite{quark1,conventional}, four-quark state \cite{4quark}, or
molecular state since it is just above the threshold of $D_s\pi$
and $DK$ \cite{molecular}, $etc.$.

In the family of excited heavy-light states in the conventional
quark model, $B_{s0}$, $B'_{s1}$, $B_{s1}$ and $B_{s2}$ are the
orbitally excited states of $B_s$ and they are the $s\bar
b~(B^*_{sJ})$ system. We know little about them since only two
candidates of them are observed in experiments. The CDF
collaboration reported their observations, $B_{s1}$ with mass
$M(B_{s1})=5829.4\pm0.7$ MeV and $B_{s2}$ with mass
$M(B_{s2})=5839.6\pm0.7$ MeV in 2008 \cite{cdf}. Later the D0
Collaboration confirmed the existence of $B_{s2}(5840)$ with mass
$M(B_{s2})=5839.6\pm1.1\pm0.7$ MeV and
 indicated that $B_{s1}(5830)$ was not observed with
 available data \cite{d0}.

Different from the lack of data in experiments, there are a lot of
theoretical efforts to investigate the properties of the
$B^{*}_{sJ}$ system. For example, the mass spectroscopy had been
estimated by the model of HQET \cite{hqet}, relativistic
constituent quark models \cite{ebert,quark,quark1,quark2,mass} and
lattice QCD \cite{qcd}. The strong decays of $B_{s0}$, $B'_{s1}$,
$B_{s1}$ and $B_{s2}$ are also studied by many authors, these
studies  helped us not only to find another two states $B_{s0}$
and $B'_{s1}$, but also to estimate the full decay widths of these
$B^{*}_{sJ}$ states.

There are large discrepancies between the existing results of the
different models, which are shown in the section of numerical
results. More careful study is needed, especially in the
relativistic models, because the relativistic corrections are
large for excited states. In this Letter, we will study the strong
decays of $B_{s0}$, $B'_{s1}$, $B_{s1}$ and $B_{s2}$ by the
improved Bethe-Salpeter(B-S) approach which is a relativistic
method based on a relativistic four-dimensional wave equation
\cite{BS,Salp}. In this model, the $B^{*}_{sJ}$ are bound states
composed of quark $s$ and anti-quark $\bar b$, with an angular
momentum $L=1$, so they are orbitally excited states and also
called $P$ wave states. The quantum numbers $J^P$ of these $P$
wave states are $0^+$ ($B_{s0}$), $1^+$ ($B'_{s1}$), $1^+$
($B_{s1}$) and $2^+$ ($B_{s2}$), the allowed strong decay modes
are $0^+ \to 0^-0^-$, $1^+ \to 1^-0^-$, $2^+ \to 0^-0^-$ and $2^+
\to 1^-0^-$, while other strong decays of $P$ wave in the final
state are ruled out by the kinematic possible mass region. For the
same reason, we have checked that in the final states of the
allowed strong decays the pseudoscalar $0^-$ state must be the
light meson ($K, \pi$), and the other one is a heavy meson ($B,
B^*, B_s, B_s^*$). Using the reduction formula, PCAC relation and
low energy theorem, we got the strong decay amplitude \cite{wang},
which is a function of the transition matrix element between two
heavy mesons. We will adopt this method to calculate the
transition matrix element by the improved B-S method in this
Letter.

Similar to the $c\bar s$ system, the $s\bar b$ (or $b\bar s$) system
is the bound state composed of a heavy quark and a light quark. Since
the heavy quark $\bar b$ is much heavier than the light quark $s$,
the heavy-light mesons can be characterized by the spin of heavy
quark
 $s_{_Q}$, the total angular momentum of light quark
$j_q=s_q+L$, and the total angular momentum $J=s_{_Q}+j_q$. For
$L=0$, the $j^P_q=\frac{1}{2}^-$ $H$ doublet, there are two states
with $J^P=0^-, 1^-$;  for $L=1$, there are two degenerate
doublets: $j^P_q=\frac{1}{2}^+$ $S$ doublet and
$j^P_q=\frac{3}{2}^+$ $T$ doublet, with the corresponding quantum
numbers $J^P=0^+, 1^+$ and $J^P=1^+, 2^+$, respectively.
 $B_{s0}$ and $B'_{s1}$ are $S$ doublet which are still missing in experiments,
  $B_{s1}(5830)$ and
$B_{s2}(5840)$ are $T$ doublet and have been observed in
experiments. Obviously, there are two $1^+$ states: $B'_{s1}$ and
$B_{s1}(5830)$, we use the notations ${\frac{1}{2}}^+$ and
${\frac{3}{2}}^+$ to describe them respectively.

Recently, we have resolved the instantaneous Bethe-Salpeter
equation, which is also called Salpeter equation, and obtained
numerical relativistic wave functions for different $J^{P(C)}$
states \cite{mass,w1}. We also give an improved formula of the
transition matrix element which is based on the Mandelstam
formulism and the relativistic Salpeter wave functions. The
corresponding transition matrix element is valid for any recoil
momentum whenever it is large or small, and we have proven that
this transition matrix element is gauge invariant when it is
necessary \cite{w4}. So in this Letter, we will use the improved
B-S method to calculate the strong decays of the orbitally excited
heavy-light states $B_{s0}$, $B'_{s1}$, $B_{s1}$ and $B_{s2}$.
According to the estimated masses theoretically, $B_{s0}$ and
$B'_{s1}$ have small masses,  we calculate the isospin symmetry
violating decay processes $B_{s0}\to B_s \pi$ and $B'_{s1}\to
B_s^* \pi$ through $\pi^0-\eta$ mixing and get small widths.
Considering the uncertainties of the masses, for $B_{s0}$ and
$B'_{s1}$, we also calculate the strong decay channels: $B_{s0}\to
B\bar K$ and $B'_{s1}\to B^*\bar K$. For $B_{s1}$ and $B_{s2}$, as
they have higher masses, the Okubo-Zweig-Iizuka (OZI) rule allowed
decays $B_{s1}\to B^*\bar K$, $B_{s2}\to B\bar K$ and $B_{s2}\to
B^*\bar K$ are permitted, in fact $B_{s1}$ and $B_{s2}$ are
observed through these decay channels.

The Letter is organized as follows. In Sec.~II, we introduce the
Bethe-Salpeter equation and the Salpeter equation. We show the
corresponding wave functions which can be obtained by solving the
Salpeter equation in Sec.~III. The method for calculating the
transition matrix elements of corresponding decays is shown in
Sec.~IV; Sec.~V show the formulations of the decay widths. Then we
show our numerical results and discussions in Sec.~VI.

\begin{figure}[htpb]
\centering
\includegraphics[width = 0.5\textwidth]{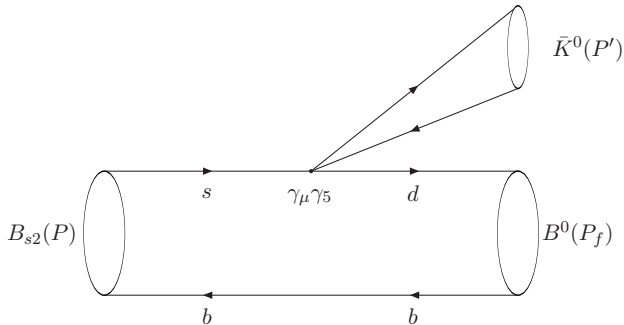}
\caption{Strong decay of $B_{s2}\to B^0\bar K^0$}
\end{figure}

\section{Instantaneous Bethe-Salpeter Equation}

In this section, we briefly review the Bethe-Salpeter equation and
its instantaneous one, the Salpeter equation, and we introduce our
notations.

The Bethe-Salpeter (BS) equation is read as \cite{BS}:
\begin{equation}
(\not\!{p_{1}}-m_{1})\chi(q)(\not\!{p_{2}}+m_{2})=
i\int\frac{d^{4}k}{(2\pi)^{4}}V(P,k,q)\chi(k)\;, \label{eq1}
\end{equation}
where $\chi(q)$ is the BS wave function, $V(P,k,q)$ is the
interaction kernel between the quark and anti-quark, and $p_{1},
p_{2}$ are the momentum of the quark 1 and anti-quark 2. The total
momentum $P$ and the relative momentum $q$ are defined as:
$$p_{1}={\alpha}_{1}P+q, \;\; {\alpha}_{1}=\frac{m_{1}}{m_{1}+m_{2}}~,$$
$$p_{2}={\alpha}_{2}P-q, \;\; {\alpha}_{2}=\frac{m_{2}}{m_{1}+m_{2}}~.$$

We divide the relative momentum $q$ into two parts,
$q_{\parallel}$ and $q_{\perp}$,
$$q^{\mu}=q^{\mu}_{\parallel}+q^{\mu}_{\perp}\;,$$
$$q^{\mu}_{\parallel}\equiv (P\cdot q/M^{2})P^{\mu}\;,\;\;\;
q^{\mu}_{\perp}\equiv q^{\mu}-q^{\mu}_{\parallel}\;.$$
Correspondingly, we have two Lorentz invariant variables:
\begin{center}
$q_{p}=\frac{(P\cdot q)}{M}\;, \;\;\;\;\;
q_{_T}=\sqrt{q^{2}_{p}-q^{2}}=\sqrt{-q^{2}_{\perp}}\;.$
\end{center}
When $\stackrel{\rightarrow}{P}=0$, they turn to the usual
component $q_{0}$ and $|\vec q|$, respectively.

In instantaneous approach, the kernel $V(P,k,q)$ takes the simple
form \cite{Salp}:
$$V(P,k,q) \Rightarrow V(k_{\perp},q_{\perp})\;.$$

Let us introduce the notations $\varphi_{p}(q^{\mu}_{\perp})$ and
$\eta(q^{\mu}_{\perp})$ for three dimensional wave function as
follows:
$$
\varphi_{p}(q^{\mu}_{\perp})\equiv i\int
\frac{dq_{p}}{2\pi}\chi(q^{\mu}_{\parallel},q^{\mu}_{\perp})\;,
$$
\begin{equation}
\eta(q^{\mu}_{\perp})\equiv\int\frac{dk_{\perp}}{(2\pi)^{3}}
V(k_{\perp},q_{\perp})\varphi_{p}(k^{\mu}_{\perp})\;. \label{eq5}
\end{equation}
Then the BS equation can be rewritten as:
\begin{equation}
\chi(q_{\parallel},q_{\perp})=S_{1}(p_{1})\eta(q_{\perp})S_{2}(p_{2})\;.
\label{eq6}
\end{equation}
The propagators of the two constituents can be decomposed as:
\begin{equation}
S_{i}(p_{i})=\frac{\Lambda^{+}_{ip}(q_{\perp})}{J(i)q_{p}
+\alpha_{i}M-\omega_{i}+i\epsilon}+
\frac{\Lambda^{-}_{ip}(q_{\perp})}{J(i)q_{p}+\alpha_{i}M+\omega_{i}-i\epsilon}\;,
\label{eq7}
\end{equation}
with
\begin{equation}
\omega_{i}=\sqrt{m_{i}^{2}+q^{2}_{_T}}\;,\;\;\;
\Lambda^{\pm}_{ip}(q_{\perp})= \frac{1}{2\omega_{ip}}\left[
\frac{\not\!{P}}{M}\omega_{i}\pm
J(i)(m_{i}+{\not\!q}_{\perp})\right]\;, \label{eq8}
\end{equation}
where $i=1, 2$ for quark and anti-quark, respectively,
 and
$J(i)=(-1)^{i+1}$. Here $\Lambda^{\pm}_{ip}(q_{\perp})$ satisfy
the relations:
\begin{equation}
\Lambda^{+}_{ip}(q_{\perp})+\Lambda^{-}_{ip}(q_{\perp})=\frac{\not\!{P}}{M}~,\;\;
\Lambda^{\pm}_{ip}(q_{\perp})\frac{\not\!{P}}{M}
\Lambda^{\pm}_{ip}(q_{\perp})=\Lambda^{\pm}_{ip}(q_{\perp})~,\;\;
\Lambda^{\pm}_{ip}(q_{\perp})\frac{\not\!{P}}{M}
\Lambda^{\mp}_{ip}(q_{\perp})=0~. \label{eq9}
\end{equation}

Introducing the notations $\varphi^{\pm\pm}_{p}(q_{\perp})$ as:
\begin{equation}
\varphi^{\pm\pm}_{p}(q_{\perp})\equiv
\Lambda^{\pm}_{1p}(q_{\perp})
\frac{\not\!{P}}{M}\varphi_{p}(q_{\perp}) \frac{\not\!{P}}{M}
\Lambda^{{\pm}}_{2p}(q_{\perp})\;, \label{eq10}
\end{equation}
and we have
$$
\varphi_{p}(q_{\perp})=\varphi^{++}_{p}(q_{\perp})+
\varphi^{+-}_{p}(q_{\perp})+\varphi^{-+}_{p}(q_{\perp})
+\varphi^{--}_{p}(q_{\perp}).
$$
Using contour integration over $q_{p}$ on both sides of
Eq.~(\ref{eq6}), we obtain:
$$
\varphi_{p}(q_{\perp})=\frac{
\Lambda^{+}_{1p}(q_{\perp})\eta_{p}(q_{\perp})\Lambda^{+}_{2p}(q_{\perp})}
{(M-\omega_{1}-\omega_{2})}- \frac{
\Lambda^{-}_{1p}(q_{\perp})\eta_{p}(q_{\perp})\Lambda^{-}_{2p}(q_{\perp})}
{(M+\omega_{1}+\omega_{2})}\;,
$$
and the full Salpeter equation:
$$
(M-\omega_{1}-\omega_{2})\varphi^{++}_{p}(q_{\perp})=
\Lambda^{+}_{1p}(q_{\perp})\eta_{p}(q_{\perp})\Lambda^{+}_{2p}(q_{\perp})\;,
$$
$$(M+\omega_{1}+\omega_{2})\varphi^{--}_{p}(q_{\perp})=-
\Lambda^{-}_{1p}(q_{\perp})\eta_{p}(q_{\perp})\Lambda^{-}_{2p}(q_{\perp})\;,$$
\begin{equation}
\varphi^{+-}_{p}(q_{\perp})=\varphi^{-+}_{p}(q_{\perp})=0\;.
\label{eq11}
\end{equation}

For the different $J^{PC}$ states, we give the general form of the
wave functions (we will talk about them in Sec.~III). Using the
last two equations in Eq.~(\ref{eq11}), we can reduce the wave
functions, then solve the wave functions by the first and second
equations in Eq.~(\ref{eq11}) to get the wave functions and mass
spectrum. We have discussed the solution of the Salpeter equation
in detail in Refs.~\cite{mass,w1}.

The normalization condition for BS wave function is:
\begin{equation}
\int\frac{q_{_T}^2dq_{_T}}{2{\pi}^2}Tr\left[\overline\varphi^{++}
\frac{{/}\!\!\!
{P}}{M}\varphi^{++}\frac{{/}\!\!\!{P}}{M}-\overline\varphi^{--}
\frac{{/}\!\!\! {P}}{M}\varphi^{--}\frac{{/}\!\!\!
{P}}{M}\right]=2P_{0}\;. \label{eq12}
\end{equation}

 In our model, Cornell
potential, a linear scalar interaction plus a vector interaction
is chosen as the instantaneous interaction kernel $V$:
\begin{equation}
V(r)=V_s(r)+V_0+\gamma_{_0}\otimes\gamma^0 V_v(r)= \lambda
r+V_0-\gamma_{_0}\otimes\gamma^0\frac{4}{3}\frac{\alpha_s}{r}~,
\end{equation}\label{vr}
 where $\lambda$ is the string constant and $\alpha_s(\vec
q)$ is the running coupling constant. In order to fit the data of
heavy quarkonia, a constant $V_0$ is often added to the scalar
confining potential. We see that $V_v(r)$ diverges at $r=0$, in
order to avoid the divergence, a factor $e^{-\alpha r}$ is added:
\begin{equation}
V_s(r)=\frac{\lambda}{\alpha}(1-e^{-\alpha r})~,
~~V_v(r)=-\frac{4}{3}\frac{\alpha_s}{r}e^{-\alpha r}~.
\end{equation}\label{vsvv}
It is easy to show that when $\alpha r\ll1$, the potential becomes
the original one. In the momentum space and the rest frame of the
bound state, the potential reads:
$$V(\vec q)=V_s(\vec q)
+\gamma_{_0}\otimes\gamma^0 V_v(\vec q)~,$$
\begin{equation}
V_s(\vec q)=-(\frac{\lambda}{\alpha}+V_0) \delta^3(\vec
q)+\frac{\lambda}{\pi^2} \frac{1}{{(\vec q}^2+{\alpha}^2)^2}~,
~~V_v(\vec q)=-\frac{2}{3{\pi}^2}\frac{\alpha_s( \vec q)}{{(\vec
q}^2+{\alpha}^2)}~,\label{eq16}
\end{equation}
where the running coupling constant $\alpha_s(\vec q)$ is chosen
as:
$$\alpha_s(\vec q)=\frac{12\pi}{33-2N_f}\frac{1}
{\log (a+\frac{{\vec q}^2}{\Lambda^{2}_{QCD}})}~.$$ With this
equation and parameters shown in Sec.~VI, one can find that
$\alpha_s(m_c)\simeq0.38$ ($N_f=3$), $\alpha_s(m_b)\simeq0.26$
($N_f=4$), and $N_f=3$ is chosen for $\bar bq$ system in this
Letter. The constants $\lambda$, $\alpha$, $V_0$ and
$\Lambda_{QCD}$ are the parameters that characterize the
potential.

\section{Relativistic Wave Functions}
In this section, by analyzing the parity and possible charge
conjugation parity of corresponding bound states, we give the
formulas of the wave functions that are in relativistic forms with
definite parity and possible charge conjugation parity symmetry.

\subsection{Wave Function for $^1S_0$ ($0^-$) state}
The general form for the relativistic wave function of a
pseudoscalar meson with the quantum number $J^P=0^-$ (or
$J^{PC}=0^{-+}$ for an equal-mass system, a $q\bar q$ quarkonium)
can be generally written as eight terms, which are constructed by
$P,\ q_{_{P_\bot}}$ and gamma matrices, because of the
instantaneous approximation, four terms with $P\cdot
q_{_{P_\bot}}$ become zero, the general form for the relativistic
Salpeter wave function of a pseudoscalar state $J^P=0^{-}$ (or
$J^{PC}=0^{-+}$) can be written as \cite{w1}:
\begin{eqnarray}\label{aa01}
\varphi_{0^-}(q_{_{P_\bot}})&=&\Big[f_1(q_{_{P_\bot}}){\not\!P}+f_2(q_{_{P_\bot}})M+
f_3(q_{_{P_\bot}})\not\!{q_{_{P_\bot}}}+f_4(q_{_{P_\bot}})\frac{{\not\!P}\not\!{q_{_{P_\bot}}}}{M}\Big]\gamma_5,
\end{eqnarray}
where $M$ is the mass of the pseudoscalar meson, and
$f_i(q_{_{P_\bot}})$ are functions of $-q^2_{_{P_\bot}}$. Due to
the last two equations of Eq.~(\ref{eq11}):
$\varphi_{0^-}^{+-}=\varphi_{0^-}^{-+}=0$, we have:
\begin{eqnarray}\label{constrain}
f_3(q_{_{P_\bot}})&=&\frac{f_2(q_{_{P_\bot}})
M(-\omega_1+\omega_2)}{m_2\omega_1+m_1\omega_2},~~~
f_4(q_{_{P_\bot}})=-\frac{f_1(q_{_{P_\bot}})
M(\omega_1+\omega_2)}{m_2\omega_1+m_1\omega_2}.
\end{eqnarray}
Then there are only two independent unknown wave functions
$f_1(q_{_{P_\bot}})$ and $f_2(q_{_{P_\bot}})$ in
Eq.~(\ref{aa01}):
\begin{eqnarray}\label{aa012}
\varphi_{0^-}(q_{_{P_\bot}})&=&\Big[f_1(q_{_{P_\bot}}){\not\!P}
+f_2(q_{_{P_\bot}})M-f_2(q_{_{P_\bot}}){\not\!{q_{_{P_\bot}}}}
\frac{M(\omega_1-\omega_2)}{m_2\omega_1+m_1\omega_2}\nonumber\\
&&+f_1(q_{_{P_\bot}}){\not\!{q_{_{P_\bot}}}\not\!P}
\frac{\omega_1+\omega_2}{m_2\omega_1 +m_1\omega_2}\Big]\gamma_5.
\end{eqnarray}
The numerical values of radial wave functions $f_1$, $f_2$ and
eigenvalue $M$ can be obtained by solving the first two equations
of Salpeter Eq.~(\ref{eq11}).

One can check that in Eq.~(\ref{aa01}), which we wrote as the wave
function for $J^P=0^-$ (or $J^{PC}=0^{-+}$) state, all the terms
except the one with $f_3$ have positive charge conjugate parity,
while $f_3$ term has negative charge conjugate parity. When we
consider the constraint relations, for equal mass system,
$\omega_1=\omega_2$, so $f_3=0$ (Eq.~(\ref{constrain})), then the
whole wave function has positive charge conjugate parity, that is
$0^{-+}$ state.

In our calculation, we obtain the numerical values of wave functions
in the center-of-mass system of the bound state, so
$q_{\parallel}$ and $q_{\perp}$ turn into the usual components
$(q_0, {\vec 0})$ and $(0,{\vec q})$, $\omega_{1}=(m_{1}^{2}+{\vec
q}^{2})^{1/2}$ and $\omega_{2}=(m_{2}^{2}+{\vec q}^{2})^{1/2}$. Then
the normalization condition reads:
\begin{eqnarray}
&\displaystyle
\int\frac{d\vec{q}}{(2\pi)^3}4f_1f_2M^2\Big\{\frac{m_1+m_2}{\omega_1
+\omega_2}+\frac{\omega_1+\omega_2}{m_1+m_2}\displaystyle
+\frac{2\vec{q}^2(m_1\omega_1+m_2\omega_2)}
{(m_2\omega_1+m_1\omega_2)^2}\Big\}=2M.
\end{eqnarray}

The numerical values of the right sides of the first two equations
in Eq.~(\ref{eq11}) are comparable, but since
$M-\omega_{1}-\omega_{2}\ll M+\omega_{1}+\omega_{2}$ for bound
state, we know that the numerical value of
${\varphi}^{++}(\vec{q})$ is much larger than that of
${\varphi}^{--}(\vec{q})$. So in the past, authors made a further
approximation of the Salpeter equation, deleting the others in
Eq.~(\ref{eq11}) except the first equation
 which is about the positive wave function
${\varphi}^{++}(\vec{q})$. This seems a reasonable approximation
since ${\varphi}^{++}(\vec{q})$ is dominant, but we point out
that, we can delete the term of ${\varphi}^{--}(\vec{q})$, but
that should be done after we solve the full Salpeter equation,
otherwise we obtain a non-relativistic wave function, not a
relativistic one. Since with the further approximation, only one
equation is left, then only one unknown $f_i$ wave function can be
solved, we have to choose $f_3=f_4=0$, and $f_1=f_2$ in
Eq.~(\ref{aa01}), then the wave function Eq.~(\ref{aa01}) becomes
$\varphi_{0^-}=f_1({\not\!P}+M)\gamma_5$, this is well known
Schrodinger wave function for a pseudoscalar. So in our
calculation, we solve the full Salpeter equations
Eq.~(\ref{eq11}), not only the first one in Eq.~(\ref{eq11}).

According to the Eq.~(\ref{eq10}) the relativistic positive wave
function of pseudoscalar $^1S_0$ state ($B$ or $B_s$ in this
Letter) in the center of mass system can be written as \cite{w1}:
\begin{eqnarray}\label{positive0-}
{\varphi}^{++}_{0^-}(\vec{q})=b_1
\left[b_2+\frac{\not\!{P}}{M}+b_3\not\!{q_{\bot}}
+b_4\frac{\not\!{q_{\bot}}\not\!{P}}{M}\right]{\gamma}_5,
\end{eqnarray}
where the $b_i$ ($i=1,~2,~3,~4$) are related to the original radial
wave function $f_i$, quark mass $m_i$, quark energy $w_i$ ($i=1,~2$)
and meson mass $M$:
$$b_1=\frac{M}{2}\left({f}_{1}(\vec{q})
+{f}_{2}(\vec{q})\frac{m_1+m_2}{w_1+w_2}\right),
b_2=\frac{w_1+w_2}{m_1+m_2}, b_3=-\frac{(m_1-m_2)}{m_1w_2+m_2w_1},
b_4=\frac{(w_1+w_2)}{(m_1w_2+m_2w_1)}.$$

Inserting the expressions of ${\varphi}^{++}_{0^-}(\vec{q})$ in
Eq.~(\ref{positive0-}) and corresponding
${\varphi}^{--}_{0^-}(\vec{q})$ (which can be easily obtained by
${\varphi}^{--}_{0^-}={\varphi}_{0^-}-{\varphi}^{++}_{0^-}$) into
the first two equations of Eq.~(\ref{eq11}), we get two coupled
integral equations (the explicit expressions can be found in
Eqs.~(24-25) in Ref.~\cite{w1}). By solving them, we obtained the
numerical values of wave functions $f_1$, $f_2$ and eigenvalue
$M$.

\subsection{Wave Function for $^3S_1$ ($1^-$) state}

Because of the instantaneous approximation, instead of 16 terms,
the general form for the relativistic wave function of vector
state $J^P=1^{-}$ (or $J^{PC}=1^{--}$ for quarkonium) can be
written as eight terms, which are constructed by $P$, $q$,
$\varepsilon$ and gamma matrices \cite{glwang}:
$$\varphi_{1^{-}}(q_{\perp})=
q_{\perp}\cdot{\varepsilon}^{(\lambda)}
\left[f_1(q_{\perp})+\frac{\not\!P}{M}f_2(q_{\perp})+
\frac{{\not\!q}_{\perp}}{M}f_3(q_{\perp})+\frac{{\not\!P}
{\not\!q}_{\perp}}{M^2} f_4(q_{\perp})\right]+
M{\not\!\varepsilon}^{(\lambda)}f_5(q_{\perp})$$
\begin{equation}+
{\not\!\varepsilon}^{(\lambda)}{\not\!P}f_6(q_{\perp})+
({\not\!q}_{\perp}{\not\!\varepsilon}^{(\lambda)}-
q_{\perp}\cdot{\varepsilon}^{(\lambda)})
f_7(q_{\perp})+\frac{1}{M}({\not\!P}{\not\!\varepsilon}^{(\lambda)}
{\not\!q}_{\perp}-{\not\!P}q_{\perp}\cdot{\varepsilon}^{(\lambda)})
f_8(q_{\perp}),\label{eq13}
\end{equation}
where ${\varepsilon}^{(\lambda)}$ is the polarization vector of the
vector meson. One should note that we use the same notations of the
radial wave functions $f_i$ for pseudoscalar and vector mesons, but
they are different. It should be indicated that we will use them for
other states (see below), but we remind the readers that their
numerical values are different for the different states.

The equations $
\varphi^{+-}_{1^{-}}(q_{\perp})=\varphi^{-+}_{1^{-}}(q_{\perp})=0\;
$ give the constraints on the components of the wave function
$\varphi_{1^{-}}(q_{\perp})$, so we have the relations
$$f_1(q_{\perp})=\frac{\left[q_{\perp}^2 f_3(q_{\perp})+M^2f_5(q_{\perp})
\right] (m_1m_2-\omega_1\omega_2+q_{\perp}^2)}
{M(m_1+m_2)q_{\perp}^2},~~~f_7(q_{\perp})=\frac{f_5(q_{\perp})M(-\omega_1+\omega_2)}
{(m_1\omega_2+m_2\omega_1)},$$
$$f_2(q_{\perp})=\frac{\left[-q_{\perp}^2 f_4(q_{\perp})+M^2f_6(q_{\perp})\right]
(m_1\omega_2-m_2\omega_1)}
{M(\omega_1+\omega_2)q_{\perp}^2},~~~f_8(q_{\perp})=\frac{f_6(q_{\perp})M(\omega_1\omega_2-m_1m_2-q_{\perp}^2)}
{(m_1+m_2)q_{\perp}^2}.$$ Then there are only four independent
wave functions $f_3(q_{\perp})$, $f_4(q_{\perp})$,
$f_5(q_{\perp})$ and $f_6(q_{\perp})$ in Eq.~(\ref{eq13}).

One can check that in Eq.~(\ref{eq13}), all the terms except those
 with $f_2$ and $f_7$ are negative under charge conjugation
operation, while the terms with $f_2$ and $f_7$ are positive.
Applying the constraint relations, for equal mass system, we found
the terms with $f_2$ and $f_7$ disappear, then the whole wave
function has negative charge conjugate parity, that is $1^{--}$
state. The similar relations hold for the following $P$ wave
states, so we will not show the details again.

 Wave functions
$f_3({\vec q})$, $f_4({\vec q})$, $f_5({\vec q})$ and $f_6({\vec
q})$ will fulfill the normalization condition:
 \begin{equation}
\int \frac{d{\vec q}}{(2\pi)^3}\frac{16\omega_1\omega_2}{3}\left\{
3f_5f_6\frac{M^2}{m_1\omega_2+m_2\omega_1}+\frac{\omega_1\omega_2-m_1m_2+{\vec
q}^2}{(m_1+m_2)(\omega_1+\omega_2)}\left[
f_4f_5-f_3\left(f_4\frac{{\vec q}^2}{M^2}+f_6\right)\right]
\right\}=2M.
 \end{equation}

The relativistic positive wave function of $^3S_1$ state ($B^*$ or
$B_s^*$ in this Letter) can be written as:
\begin{eqnarray}
{\varphi}_{1^-}^{++}(\vec{q})&=&b_1\not\!{\varepsilon}^{(\lambda)}+b_2\not\!{\varepsilon}^{(\lambda)}\not\!{P}
+b_3(\not\!{q_{\bot}}\not\!{\varepsilon}^{(\lambda)}-q_{\bot}\cdot{\varepsilon}^{(\lambda)})
+b_4(\not\!{P}\not\!{\varepsilon}^{(\lambda)}\not\!{q_{\bot}}-\not\!{P}q_{\bot}\cdot{\varepsilon}^{(\lambda)})
\nonumber\\
&&+q_{\bot}\cdot{\varepsilon}^{(\lambda)}(b_5+b_6\not\!{P}+b_7\not\!{q_{\bot}}+b_8\not\!{q_{\bot}}\not\!{P}),
\end{eqnarray}
where we first defined the parameters $n_i$ which are functions of
$f_i$ ($^3S_1$ wave functions):
$$n_1={f}_{5}(\vec{q})
-{f}_{6}(\vec{q})\frac{(w_1+w_2)}{(m_1+m_2)}, n_2={f}_{5}(\vec{q})
-{f}_{6}(\vec{q})\frac{(m_1+m_2)}{(w_1+w_2)}, n_3={f}_{3}(\vec{q})
+{f}_{4}(\vec{q})\frac{(m_1+m_2)}{(w_1+w_2)},$$ then we defined
the parameters $b_i$ which are functions of $f_i$ and $n_i$:
$$b_1=\frac{M}{2}n_1, b_2=-\frac{M}{2}\frac{(m_1+m_2)}{(w_1+w_2)}n_1,
 b_3=\frac{M}{2}\frac{(w_2-w_1)}{(m_1w_2+m_2w_1)}n_1, b_4=\frac{1}{2}\frac{(w_1+w_2)}{(w_1w_2+m_1m_2-{q_{\bot}^{2}})}n_1,$$
 $$b_5=\frac{1}{2M}\frac{m_1+m_2}{(w_1w_2+m_1m_2+{q_{\bot}^{2}})}(M^2n_2+{q_{\bot}^{2}}n_3),
  b_6=\frac{1}{2M^2}\frac{w_1-w_2}{(w_1w_2+m_1m_2+{q_{\bot}^{2}})}(M^2n_2+{q_{\bot}^{2}}n_3),$$
$$b_7=\frac{n_3}{2M}-\frac{f_6(\vec{q})M}{(m_1w_2+m_2w_1)},
 b_8=\frac{1}{2M^2}\frac{w_1+w_2}{m_1+m_2}n_3-f_5(\vec{q})\frac{w_1+w_2}{(m_1+m_2)(w_1w_2+m_1m_2-{q_{\bot}^{2}})}.$$

Similar to the method in last subsection, where we obtained the
wave functions and eigenvalues for pseudoscalar states, inserting
${\varphi}_{1^-}^{++}(\vec{q})$ and corresponding
${\varphi}_{1^-}^{--}(\vec{q})$ into the first two equations of
Eq.~(\ref{eq11}), we obtained four independent coupled integral
equations (Eqs.~(37-40) in Ref.~\cite{mass}), by solving them, we
obtained the numerical results of mass spectra and wave functions.
For other states, see below, we will not show the details of how
to solve them, interested reader can find the details elsewhere,
for example, in Ref.~\cite{mass}.

One should also note that we cite same notations of $m_i$, $w_i$,
$M$, $P$, and $b_i$ as used in last subsection for pseudoscalar
meson, but they are different for different states. And we also
use the same notations for other mesons, like the following $P$
wave states.

\subsection{ Wave function for $^3P_0$ state}

The general form for the relativistic Salpeter wave function of
$^3P_0$ state, which $J^P=0^+$ (or $J^{PC}=0^{++}$ for equal mass
system), can be written as \cite{w2}:
\begin{equation}\varphi_{0^{+}}(q_{\perp})=
f_1(q_{\perp}){\not\!q}+f_2(q_{\perp})\frac{{\not\!P}
{\not\!q}_{\perp}}{M} +f_3(q_{\perp})M+f_4(q_{\perp}){\not\!P}.
\end{equation}
 The equations
$\varphi^{+-}_{0^{+}}(q_{\perp})=\varphi^{-+}_{0^{+}}(q_{\perp})=0\;
$ give the constraints on the components of the wave functions, so
we have the relations
$$f_3(q_{\perp})=\frac{f_1(q_{\perp})q_{\perp}^2(m_1+m_2)}
{M(\omega_1\omega_2+m_1m_2+q_{\perp}^2)},~~~
f_4(q_{\perp})=\frac{f_2(q_{\perp})q_{\perp}^2(\omega_1-\omega_2)}
{M(m_1\omega_2+m_2\omega_1)}.$$ Then there are only two
independent wave functions $f_1(q_{\perp})$ and $f_2(q_{\perp})$.
From Eq.~(\ref{eq11}), we obtain two coupled integral equations,
by solving them, we obtain the numerical results of mass spectra
and wave functions.

The normalization condition for the $^3P_0$ wave function is:
 \begin{equation}
\int \frac{d{\vec q}}{(2\pi)^3}\frac{16f_1f_2
\omega_1\omega_2{\vec q}^2}{m_1\omega_2+m_2\omega_1}=2M.
 \end{equation}

The relativistic positive energy wave function of $B_{s0}~(^3P_0)$
can be written as:
\begin{equation}
{\varphi}_{0^+}^{++}(\vec{q})=a_1(\not\!{q_{\bot}}+a_2\frac{\not\!{P}\not\!{q_{\bot}}}{M}
+a_3+a_4\frac{\not\!{P}}{M}),
\end{equation}
where the parameters $a_i$ are functions of $f_1$ and $f_2$ ($0^+$
wave function) and are defined as:
$$a_1=\frac{1}{2}\left({f}_{1}(\vec{q})
+{f}_{2}(\vec{q})\frac{m_1+m_2}{w_1+w_2}\right),
a_2=\frac{w_1+w_2}{m_1+m_2},
a_3=q_{\bot}^2\frac{(w_1+w_2)}{m_1w_2+m_2w_1},
a_4=\frac{(m_2w_1-m_1w_2)}{(m_1+m_2)}.$$

\subsection{Wave function for $^3P_1$ state}

The general form for the Salpeter wave function of $^3P_1$ state,
which $J^P=1^+$ (or $J^{PC}=1^{++}$ for equal mass system), can be
written as \cite{w2}:
\begin{equation}\varphi_{^3P_1}(q_{\perp})=i\epsilon_{\mu\nu\alpha\beta}
P^{\nu}q_{\perp}^{\alpha}\varepsilon^{\beta}\left[f_1M\gamma^{\mu}+
f_2{\not\!P}\gamma^{\mu}+f_3{\not\!q}_{\perp}\gamma^{\mu}
+if_4\epsilon^{\mu\rho\sigma\delta}
q_{\perp\rho}P_{\sigma}\gamma_{\delta}\gamma_{5}/M \right]/M^2,
\end{equation}
where ${\varepsilon}^{(\lambda)}$ is the polarization vector of
the $^3P_1$ state.

The constraint equations give us the relations:
$$f_3(q_{\perp})=\frac{f_1(q_{\perp})M(m_1\omega_2-m_2\omega_1)}
{q_{\perp}^2(\omega_1+\omega_2)},~~~
f_4(q_{\perp})=\frac{f_2(q_{\perp})M(-\omega_1\omega_2+m_1m_2+q_{\perp}^2)}
{q_{\perp}^2(m_1+m_2)}.$$

The normalization condition for the $^3P_1$ wave function is:
 \begin{equation}
\int \frac{d{\vec q}}{(2\pi)^3}\frac{32f_1f_2
\omega_1\omega_2(\omega_1\omega_2-m_1m_2+{\vec
q}^{2})}{3(m_1+m_2)(\omega_1+\omega_2)}=2M.
 \end{equation}

The relativistic positive energy wave function of $^3P_1$ state
can be written as:
\begin{equation}
{\varphi}_{^3P_1}^{++}(\vec{q})=i{\epsilon}_{\mu\nu\alpha\beta}P^{\nu}q^{\alpha}_{\bot}{\varepsilon}^{\beta(\lambda)}
a_1[M{\gamma}^{\mu}+a_2{\gamma}^{\mu}\not\!{P}+a_3{\gamma}^{\mu}\not\!{q_{\bot}}
+a_4{\gamma}^{\mu}\not\!{P}\not\!{q_{\bot}}]/M^2,
\end{equation}
where the parameters $a_i$ are functions of $f_1$ and $f_2$
($^3P_1$ wave function) and are defined as:
$$a_1=\frac{1}{2}\left(f_1(\vec{q})+f_2(\vec{q})\frac{w_1+w_2}{m_1+m_2}\right),
 a_2=-\frac{m_1+m_2}{w_1+w_2}, a_3=\frac{M(w_1-w_2)}{m_1w_2+m_2w_1}, a_4=-\frac{(m_1+m_2)}{m_1w_2+m_2w_1}.$$

\subsection{Wave function for $^1P_1$ state}

The general form for the Salpeter wave function of $^1P_1$ state,
which $J^P=1^+$ (or $J^{PC}=1^{+-}$ for equal mass system), can be
written as \cite{w2}:
\begin{equation}\varphi_{^1P_1}(q_{\perp})=
q_{\perp}\cdot{\varepsilon^{(\lambda)}}\left[
f_1(q_{\perp})+f_2(q_{\perp})\frac{{\not\!P}}{M}
+f_3(q_{\perp})\frac{{\not\!q_{\perp}}}{M}+
f_4(q_{\perp})\frac{\not\!{P}{\not\!q}}{M^2}\right]\gamma_5,
\end{equation}
where ${\varepsilon^{(\lambda)}}$ is the polarization vector of
the $^1P_1$ state.

The constraint equations provide us the relations:
$$f_3(q_{\perp})=-\frac{f_1(q_{\perp})M(m_1-m_2)}
{(\omega_1\omega_2+m_1m_2-q_{\perp}^2)},~~~
f_4(q_{\perp})=-\frac{f_2(q_{\perp})M(\omega_1+\omega_2)}
{(m_1\omega_2+m_2\omega_1)}.$$

The normalization condition for the $^1P_1$ wave function is:
 \begin{equation}
\int \frac{d{\vec q}}{(2\pi)^3}\frac{16f_1f_2
\omega_1\omega_2{\vec q}^2}{3(m_1\omega_2+m_2\omega_1)}=2M.
 \end{equation}

 The relativistic positive energy wave function of $^1P_1$ can be written as:
\begin{equation}
{\varphi}_{^1P_1}^{++}(\vec{q})=(\varepsilon^{(\lambda)}\cdot
q_{\bot}) a_1\left[1+a_2\frac{\not\!P}{M}
+a_3\not\!{q_{\bot}}+a_4\frac{\not\!{q_{\bot}}\not\!P}{M}\right]{\gamma}_5,
\end{equation}
where the parameters $a_i$ are functions of $f_1$ and $f_2$
($^1P_1$ wave function) and are defined as:
$$a_1=\frac{1}{2}\left(f_1(\vec{q}+f_2(\vec{q})\frac{w_1+w_2}{m_1+m_2})\right), a_2=\frac{m_1+m_2}{w_1+w_2},
a_3=-\frac{w_1-w_2}{m_2w_1+m_1w_2},
a_4=\frac{m_1+m_2}{m_2w_1+m_1w_2}.$$

The wave functions of two physical $1^+$ states (or
${\frac{1}{2}}^+$ and ${\frac{3}{2}}^+$) are the mixing of
${\varphi}_{^3P_1}^{++}(\vec{q})$ and
${\varphi}_{^1P_1}^{++}(\vec{q})$, see Eq.~(\ref{mix}) below.
\subsection{Wave function for $^3P_2$ state}

The general form for the relativistic wave function of tensor
$J^{P}=2^{+}$ state (or $J^{PC}=2^{++}$ for equal mass system) can
be written as \cite{w3}:
$$\Psi_{2^{+}}(\vec{q})=
{\varepsilon}_{\mu\nu}^{(\lambda)}{q_{\perp}^{\nu}}
\left\{{q_{\perp}^{\mu}}\left[f_1(\vec{q})+\frac{\not\!P}{M}f_2(\vec{q})+
\frac{{\not\!q}_{\perp}}{M}f_3(\vec{q})+\frac{{\not\!P}
{\not\!q}_{\perp}}{M^2} f_4(\vec{q})\right]\right.$$
\begin{equation}\left.+
{\gamma^{\mu}}\left[Mf_5(\vec{q})+ {\not\!P}f_6(\vec{q})+
{\not\!q}_{\perp} f_7(\vec{q})\right]+\frac{i}{M}
f_8(\vec{q})\epsilon^{\mu\alpha\beta\gamma}
P_{\alpha}q_{\perp\beta}\gamma_{\gamma}\gamma_{5}\right\},\label{eq3}
\end{equation}
where ${\varepsilon}_{\mu\nu}^{(\lambda)}$ is the polarization
tensor of the $2^{+}$ state. The constraint equations give further
relations:
$$f_1(\vec{q})=\frac{\left[q_{\perp}^2 f_3(\vec{q})+M^2f_5(\vec{q})
\right](\omega_1+\omega_2)-M^2f_5(\vec{q})(\omega_1-\omega_2)}
{M(m_1\omega_2+m_2\omega_1)},~$$
$$f_2(\vec{q})=\frac{\left[q_{\perp}^2 f_4(\vec{q})-M^2f_6(\vec{q})\right]
(\omega_1-\omega_2)}
{M(m_1\omega_2+m_2\omega_1)},~~$$\begin{equation}~~f_7(\vec{q})=\frac{f_5(\vec{q})M(\omega_1-\omega_2)}
{m_1\omega_2+m_2\omega_1},~~~~f_8(\vec{q})=\frac{f_6(\vec{q})M(\omega_1+\omega_2)}
{m_1\omega_2+m_2\omega_1}.\end{equation} Only four independent wave
functions $f_3(\vec{q})$, $f_4(\vec{q})$, $f_5(\vec{q})$ and
$f_6(\vec{q})$, the numerical values and the bound state mass $M$
can be obtained by solving the full Salpeter equation.

These four independent wave functions fulfil the normalization
condition:
$$\int \frac{d{\vec q}}{(2\pi)^3}\frac{16~\omega_1\omega_2~{\vec
q}^2}{15(m_1\omega_2+m_2\omega_1)}\left\{
f_5~f_6~M^2\left[5+\frac{(m_1+m_2)(m_2\omega_1-m_1\omega_2)}
{\omega_1\omega_2(\omega_1+\omega_2)}\right]\right.
$$\begin{equation}\left.
+f_4~f_5~{\vec
q}^2\left[2+\frac{(m_1+m_2)(m_2\omega_1-m_1\omega_2)}
{\omega_1\omega_2(\omega_1+\omega_2)}\right] -2~{\vec
q}^2f_3\left(f_4\frac{{\vec q}^2}{M^2}+f_6\right) \right\}=2M.
 \end{equation}

 The relativistic positive energy wave function of $B_{s2}~(^3P_2)$ can be written
as:
 \begin{eqnarray}
&&{\varphi}^{++}_{^3P_2}={\varepsilon}^{(\lambda)}_{\mu
\nu}q^{\nu}_{\bot}\{q^{\mu}_{\bot}[a_1+a_2\frac{\not\!P}{M}
+a_3\frac{{\not\!q_{\bot}}}{M}+a_4\frac{\not\!q_{\bot}\not\!P}{M^2}]
+{\gamma}^{\mu}[a_5+a_6\frac{\not\!P}{M}+a_7\frac{\not\!q_{\bot}}{M}+a_8\frac{\not\!P\not\!q_{\bot}}{M^2}]\},
\end{eqnarray}
similar to $1^-$ state, we first defined $n_i$ as:
$$n_1=f_3(\vec{q})+f_4(\vec{q})\frac{m_1+m_2}{w_1+w_2},~
n_2=f_5(\vec{q})-f_6(\vec{q})\frac{w_1+w_2}{m_1+m_2},$$ then we
defined the parameters $a_i$:
$$a_1=\frac{(w_1+w_2)q^2_{\bot}}{2M(m_1w_2+m_2w_1)}n_1+\frac{(f_5(\vec{q})w_2-f_6(\vec{q})m_2)M}{m_1w_2+m_2w_1},$$
$$a_2=\frac{(m_1-m_2)q^2_{\bot}}{2M(m_1w_2+m_2w_1)}n_1+\frac{(f_6(\vec{q})w_2-f_5(\vec{q})m_2)M}{m_1w_2+m_2w_1},$$
$$a_3=\frac{1}{2}n_1+\frac{f_6(\vec{q})M^2}{m_1w_2+m_2w_1},~
a_4=\frac{1}{2}(-\frac{w_1+w_2}{m_1+m_2})n_1+\frac{f_5(\vec{q})M^2}{m_1w_2+m_2w_1},$$
$$a_5=\frac{M}{2}n_2,~a_6=\frac{M(m_1+m_2)}{2(w_1+w_2)}n_2,~
a_7=\frac{M^2(w_1-w_2)}{2(m_2w_1+m_1w_2)}n_2,~a_8=\frac{M^2(m_1+m_2)}{2(m_2w_1+m_1w_2)}n_2.$$

\section{Transition Matrix Element}

In this section, we show the method to formulate the transition
matrix element, which is general for all the decay channels in
this Letter.

Considering the limitations of phase spaces, there are seven
dominant strong decay channels for $B^{*}_{sJ}$ states: $B_{s0}\to
B^0_s \pi^0$, $B_{s0}\to B\bar K$, $B'_{s1}\to B_s^{*0} \pi^0$,
$B'_{s1}\to B^*\bar K$, $B_{s1}\to B^*\bar K$, $B_{s2}\to B\bar K$
and $B_{s2}\to B^*\bar K$ (where $B\bar K=B^+K^-+B^0\bar{K^0}$ and
$B^*\bar K=B^{*+}K^-+B^{*0}\bar{K^0}$). Since all the light mesons
in final states are pseudoscalars, we can give a unique
formulation of the transition matrix element for these seven decay
channels. By using the reduction formula, PCAC relation and low
energy theorem, taking the channel $B_{s2}\to B\bar K$ as an
example, see Figure 1, the corresponding transition matrix element
can be written as \cite{wang}:
\begin{eqnarray}\label{t}
T(B_{s2}\to B^0\bar
K^0)=\frac{P'^{\mu}}{f_{_K}}<B^0(P_f)|\bar{d}{\gamma}_{\mu}\gamma_5s|B_{s2}(P)>,
\end{eqnarray}
where $f_{_K}$ is the decay constant of pseudoscalar $K$ meson, $P'$
is the momentum of $K$. The contribution of the light pseudoscalar
is reduced to a factor $\frac{P'^{\mu}}{f_{_K}}$, then the main part
of the calculation  in Eq.~(\ref{t}) is to calculate the transition
element $<B^0(P_f)|\bar{d}{\gamma}_{\mu}\gamma_5s|B_{s2}(P)>$.

If we further choose the instantaneous approach, according to the
Mandelstam formalism \cite{Mand}, at the leading order, the
transition matrix element can be written as an integral equation
of the corresponding initial and final state wave functions
\cite{w4}:
\begin{eqnarray}\label{previous}
\langle B^0(P_f)|\bar{d}{\gamma}_{\mu}\gamma_5s|B_{s2}(P)\rangle
=\int\frac{d\vec{q}}{(2\pi)^3}Tr
\Big\{\bar{\varphi}^{\prime++}_{B^0}(\vec{q'}){\gamma}_{\mu}{\gamma}_5{\varphi}^{++}_{B_{s2}}
(\vec{q})\frac{\not\!P}{M}\Big\}\,,
\end{eqnarray}
where $P$ and $M$ are the momentum and mass of initial state
$B_{s2}$; $\vec{q}$ and
$\vec{q'}=\vec{q}+\frac{m_b}{m_b+m_d}\times\vec{P}_f$ are the
relative momenta of quark and anti-quark in the initial state
$B_{s2}$ and the final state $B^0$, respectively, which are
defined as $\vec{q}=\vec{p}_s=-\vec{p}_b$ and
$\vec{q'}=\frac{m_b}{m_b+m_d}\times\vec{P}_f-\vec{p}_b$  in the
 center of mass system of initial state $B_{s2}$;
 ${\varphi}^{++}
(q_{\bot})$ and $\bar{\varphi}^{\prime++}(q_{\bot}^{\prime})$ are
the positive energy wave functions of $B_{s2}$ and $B^0$, which
are given in last section.

In our model, improved B-S method, which is based on the
constituent quark model, we give the forms of wave functions by
considering the quantum number $J^P$ or $J^{PC}$ for different
states, and these states in our model are labelled as $^1S_0$
($0^-$ state), $^3S_1$ ($1^-$ state), $^3P_J$ ($J=1,~2,~3$)
($0^+$, $1^+$, $2^+$) and $^1P_1$ ($1^+$).
  For the unequal mass system, the $^3P_1$ and $^1P_1$ states are not physical states,
  the two physical states ${\frac{1}{2}}^+$ and
${\frac{3}{2}}^+$, which are the mixtures of them, can be
expressed as \cite{rosner,hqet,aa,wise}:
\begin{eqnarray}\label{mix}
&&|B'_{s1}>=|{\frac{1}{2}}^+>=\sin\theta|^1P_1>-\cos\theta|^3P_1>,\nonumber\\
&&|B_{s1}>=|{\frac{3}{2}}^+>=\cos\theta|^1P_1>+\sin\theta|^3P_1>,
\end{eqnarray}
where $\theta$ is the mixing angle and $\theta\approx35.3^\circ$
in the heavy quark limit.

The strong decay amplitudes can be described by the strong
coupling constants, they are defined as:
\begin{eqnarray}
&&T(B_{s0}\to B_s\pi)=G_{B_{s0}B_s\pi},\nonumber\\
&&T(B_{s0}\to B\bar K)=G_{B_{s0}B\bar K},\nonumber\\
&&T(B'_{s1}\to
B_s^*\pi)=G_{B'_{s1}B_s^*\pi}({\varepsilon}_1^{(\lambda')}\cdot
v)({\varepsilon}^{(\lambda)}\cdot
v')+F_{B'_{s1}B_s^*\pi}({\varepsilon}_1^{(\lambda')}\cdot{\varepsilon}^{(\lambda)}),\nonumber\\
&&T(B'_{s1}\to B^{*}\bar K)=G_{B'_{s1}B^{*}\bar
K}({\varepsilon}_1^{(\lambda')}\cdot
v)({\varepsilon}^{(\lambda)}\cdot
v')+F_{B'_{s1}B^{*}\bar K}({\varepsilon}_1^{(\lambda')}\cdot{\varepsilon}^{(\lambda)}),\nonumber\\
&&T(B_{s1}\to B^{*}\bar K)=G_{B_{s1}B^{*}\bar
K}({\varepsilon}_1^{(\lambda')}\cdot
v)({\varepsilon}^{(\lambda)}\cdot
v')+F_{B_{s1}B^{*}\bar K}({\varepsilon}_1^{(\lambda')}\cdot{\varepsilon}^{(\lambda)}),\nonumber\\
&&T(B_{s2}\to
B\bar K)=G_{B_{s2}B\bar K}{\varepsilon}_{\mu\nu}^{(\lambda)}v^{\prime\mu}v^{\prime\nu},\nonumber\\
&&T(B_{s2}\to B^{*}\bar K)=iG_{B_{s2}B^{*}\bar
K}{\varepsilon}_{\mu\nu}^{(\lambda)}v^{\prime\nu}{\epsilon}^{{\varepsilon}_1^{(\lambda')}vv'\mu},
\end{eqnarray}
where $G_{B_{s0}B_s\pi},\cdots G_{B_{s2}B^{*}\bar K}$ are the
strong coupling constants; $v=\frac{P}{M}$ and
$v'=\frac{P_f}{M_f}$ are four-velocities of initial state and
final state;  ${\varepsilon}$ is the polarization vector of
initial state $B'_{s1}$ or $B_{s1}$, ${\varepsilon}_1$ is the
polarization vector of final state $B_s^*$ or $B^*$,
${\varepsilon}_{\mu\nu}$ is the polarization vector of initial
state $B_{s2}$; $P,~P_f$ are the four-momenta of initial and final
heavy states, respectively.

\section{Decay Widths}

The decay channels of $B_{s0}\to B\bar K$, $B'_{s1}\to B^*\bar K$,
$B_{s1}\to B^{*}\bar K$, $B_{s2}\to B\bar K$ and $B_{s2}\to
B^{*}\bar K$ are OZI rule allowed, through the Eq. (\ref{t}) and Eq.
(\ref{previous}) the calculations of corresponding decay widths are
straightforward:
\begin{eqnarray}
&& \Gamma_{{B_{s0}B\bar K}}=\frac{|\vec P_f|}{8\pi
M^2}\left|{T(B_{s0}\to B\bar K)}\right|^2,\ \ \
\Gamma_{{B'_{s1}B^*\bar K}}=\frac{|\vec P_f|}{24\pi
M^2}\sum_{\lambda, \lambda'}\left|{T(B'_{s1}\to
B^*\bar K)}\right|^2,\nonumber\\
&&\Gamma_{{B_{s1}B^{*}\bar K}}=\frac{|\vec P_f|}{24\pi
M^2}\sum_{\lambda, \lambda'}\left|T(B_{s1}\to B^{*}\bar
K)\right|^2,\ \ \
 \Gamma_{B_{s2}B\bar K}=\frac{|\vec P_f|}{40\pi
M^2}\sum_{\lambda}\left|T(B_{s2}\to B\bar K)\right|^2,
\nonumber\\
&& \Gamma_{B_{s2}B^{*}\bar K}=\frac{|\vec P_f|}{40\pi
M^2}\sum_{\lambda,\lambda'}\left|T(B_{s2}\to B^{*}\bar
K)\right|^2.
\end{eqnarray}

The decays of $B_{s0}\to B_s\pi$ and $B'_{s1}\to B_s^*\pi$ violate
the isospin symmetry which only occur through $\pi^0-\eta$ mixing
\cite{mix1}. According to Dashen' theorem \cite{dashen}, the decay
widths can be written as:
\begin{eqnarray}
\Gamma_{{B_{s0}B_s\pi}}=\frac{|\vec P_f|}{8\pi
M^2}\left|\frac{T(B_{s0}\to
B_s\pi)t_{\pi\eta}}{m^2_{\pi}-m^2_{\eta}}\right|^2,\ \ \
\Gamma_{{B'_{s1}B_s^*\pi}}=\frac{|\vec P_f|}{24\pi
M^2}\sum_{\lambda, \lambda'}\left|\frac{T(B'_{s1}\to
B_s^*\pi)t_{\pi\eta}}{m^2_{\pi}-m^2_{\eta}}\right|^2,
\end{eqnarray}
where $t_{\pi\eta}=<\pi^0|\mathcal{H}|\eta>$ is the ${\pi}^0-\eta$
transition matrix, $m_{\pi}$ and $m_{\eta}$ are the masses of
$\pi$ and $\eta$. The chosen value of $t_{\pi\eta}=-0.003$ GeV$^2$
\cite{dashen} is very small, which result in narrow decay widths
of these two channels.

\section{Numerical Results and Discussions}

 In order to
obtain the masses and wave functions of initial and final states,
which are used to calculate the transition matrix elements and
decay widths, we solve the instantaneous B-S equation
\cite{BS,Salp}, and the parameters are chosen as: $a=e=2.7183,
\lambda=0.21$ GeV$^2$, ${\Lambda}_{QCD}=0.27$ GeV, $\alpha=0.06$
GeV, $m_b=4.96$ GeV, $m_s=0.5$ GeV, $m_u=0.305$ GeV, $m_d=0.311$
GeV and $V_0$ for the B-S kernel as same as
Refs.~\cite{w1,glwang,w2,w3,mass}. The numerical values of these
parameters are obtained by fitting data, for examples:
$M_{B_{s1}}=5.830$ GeV, $M_{B_{s2}}=5.840$ GeV, $M_{B}=5.279$ GeV,
$M_{B^*}=5.325$ GeV, $M_{B_s}=5.366$ GeV, $M_{B_s^*}=5.415$ GeV
\cite{pdg}.

With this set of parameters, and varying all the input parameters
simultaneously within $\pm 5 \%$ of the central values, we obtain
the light doublet masses:
\begin{equation}M_{B_{s0}}=5.723\pm0.280~ {\rm GeV},~
M_{B'_{s1}}=5.774\pm0.330~ {\rm GeV}.\end{equation}  The other
input parameters in this Letter are the masses of light mesons:
$M_{\pi}=0.139$ GeV, $M_{\eta}=0.548$ GeV, $M_K=0.494$ GeV, and
the decay constants $f_{\pi}=130$ MeV, $f_{K}=156$ MeV \cite{pdg}.

With these parameters, we solve the instantaneous Salpeter
equations for different states, and obtain the masses and
relativistic wave functions. We calculate the transition matrix
elements by the wave functions numerically, and get the
strong coupling constants:
\begin{eqnarray}
&&G_{B_{s0}B_s\pi}=17.5\pm2.5~{\rm GeV}; \ \ \
G_{B'_{s1}B_s^*\pi}=-16.3\pm2.2 ~{\rm GeV},
\ \ \ F_{B'_{s1}B_s^*\pi}=17.2\pm3.3~{\rm GeV};\nonumber\\
&& G_{B_{s1}B^{*+}K^-}=-2.39\pm0.36~{\rm TeV}, \ \ \ F_{B_{s1}B^{*+}K^-}=0.320\pm0.065~{\rm GeV};\nonumber\\
&& G_{B_{s1}B^{*0}\bar K^0}=-2.48\pm0.37~{\rm TeV}, \ \ \ F_{B_{s1}B^{*0}\bar K^0}=0.227\pm0.050~{\rm GeV};\nonumber\\
&&G_{B_{s2}B^+K^-}=1.96\pm0.26~{\rm TeV}, \ \ \ \ G_{B_{s2}B^0\bar K^0}=1.97\pm0.26~{\rm TeV};\nonumber\\
&& G_{B_{s2}B^{*+}K^-}=2.33\pm0.65~{\rm TeV},\ \ \ \
G_{B_{s2}B^{*0}\bar K^0}=2.35\pm0.65~{\rm TeV}.
\end{eqnarray}

Considering the uncertainties of the masses, for mesons $B_{s0}$
and $B'_{s1}$, the upper limits of the masses are
$M_{B_{s0}}\approx6.00$ GeV and $M_{B'_{s1}}\approx6.10$ GeV,
which are above the threshold of $B\bar K$, $B^*\bar K$, so with
these upper limit values of masses, we also calculated the OZI
allowed channels:
\begin{eqnarray}
&&G_{B_{s0}B^0\bar K^0}=28.6~{\rm GeV}, \ \ G_{B_{s0}B^+K^-}=28.7~{\rm GeV};\nonumber\\
&&G_{B'_{s1}B^{*0}\bar K^0}=2.34~{\rm TeV}, \ \ F_{B'_{s1}B^{*0}\bar
K^0}=6.34~{\rm GeV};\nonumber\\
&& G_{B'_{s1}B^{*+}K^-}=2.32~{\rm TeV},\ \
F_{B'_{s1}B^{*+}K^-}=6.34~{\rm GeV}.
\end{eqnarray} The corresponding decay widths are shown in Table~I.

\begin{table}\caption{The strong decay widths of $B_{s0}$ and $B'_{s1}$ in units of MeV.
Here we only calculated the cases that the upper limit values of
masses of $B_{s0}$ and $B'_{s1}$ are chosen, and we take
$\theta=35.3^{\circ}$.}
\begin{center}
\begin{tabular}{|c|c|}
\hline \hline Mode&Ours \\
\hline
 $B_{s0}\to B^{+}{K^-}$&67.1\\
  $B_{s0}\to B^{0}{\bar K^0}$&67.0\\
  \hline
$B'_{s1}\to B^{*+}{K^-}$&37.7\\
$B'_{s1}\to B^{*0}{\bar K^0}$&37.6\\
\hline
\end{tabular}
\end{center}
\end{table}
\begin{table}\caption{The strong decay widths of $B_{s0}$ and $B'_{s1}$ in units of
keV, where the decays of $B_{s0}\to B{\bar K}$ and $B'_{s1}\to
B^*{\bar K}$ are calculated with the upper limit masses of
$B_{s0}$ and $B'_{s1}$. And the mixing angle $\theta=35.3^{\circ}$
is used.}
\begin{center}
\begin{tabular}{|c|c|c|c|c|c|c|}
\hline \hline
Mode&Ours&\cite{wangzg}&\cite{L4,L5}&\cite{qq}&\cite{16}&\cite{zhao} \\
 $B_{s0}\to B_s\pi$&$13.6\pm5.6$&6.8$\sim$30.7&1.54~\cite{L4}&55.2$\sim$89.9&21.5&--\\
 $B_{s0}\to B{\bar K}$&134 MeV&--&--&--&--&227 MeV\\
 \hline
$B'_{s1}\to B_s^*\pi$&$13.8\pm3.6$&
5.7$\sim$20.7&10.36~\cite{L5}&57.0$\sim$94.0&21.5&
\\
 $B'_{s1}\to B^*{\bar K}$&75.3 MeV&--&--&--&--&149 MeV\\
\hline \hline
\end{tabular}
\end{center}
\end{table}

\begin{table}\caption{The strong decay widths of $B_{s1}$ and $B_{s2}$ in units of MeV. Here we take $\theta=35.3^{\circ}$.}
\begin{center}
\begin{tabular}{|c|c|}
\hline \hline Mode&Ours \\
\hline
 $B_{s1}\to B^{*+}{K^-}$&$0.028\pm0.0075$\\
  $B_{s1}\to B^{*0}{\bar K^0}$&$0.013\pm0.0036$\\
  \hline
$B_{s2}\to B^+{K^-}$&$0.83\pm0.23$\\
$B_{s2}\to B^0\bar K^0$&$0.72\pm0.20$\\
\hline
 $B_{s2}\to B^{*+}{K^-}$&$0.091\pm0.052$
\\
 $B_{s2}\to B^{*0}{\bar K^0}$&$0.057\pm0.032$
\\
\hline \hline
\end{tabular}
\end{center}
\end{table}

\begin{table}\caption{The strong decay widths of $B_{s1}$ and $B_{s2}$ in units of MeV. Here we take $\theta=35.3^{\circ}$.}
\begin{center}
\begin{tabular}{|c|c|c|c|c|c|c|c|c|}
\hline \hline Mode&Ours& \cite{quark} &\cite{liu}&
\cite{supp}&\cite{zhao}&\cite{aff}&\cite{hill} \\
\hline
 $B_{s1}\to B^*\bar{K}$&$0.041\pm0.011$&--&0.098&3.5&$0.4\sim1$&0.28&$<1$\\
$B_{s2}\to B\bar{K}$&$1.55\pm0.43$&2.6(1.9)&4.6&-&2&$7\pm3$&1\\
$B_{s2}\to
B^*\bar{K}$&$0.148\pm0.084$&0.07(0.05)&0.4&3.2&0.12&--&$ <1$
\\
\hline \hline
\end{tabular}
\end{center}
\end{table}

In Table.~I, II, III and IV, we show numerical results of strong
decay widths, which are predicted by us and other authors. For
$B_{s0}$ and $B'_{s1}$, our results are consistent with the ones
presented by references \cite{wangzg} and \cite{L5}, close to the
results of reference \cite{16}, but much larger than the ones of
\cite{L4}, and much smaller than the predictions of \cite{qq}.
Though the predicted masses of the $P$-wave $B^*_{sJ}$ states are
similar for different models, the predicted decay widths are much
different. The situation is similar in other channels. For
example, our prediction of $B_{s1}\to B^*\bar K$ is close to the
result of reference \cite{liu}, and we get the narrowest decay
width.

In our calculation, we choose a small value of mixing parameter,
$t_{\pi\eta}=-0.003$ GeV$^2$, which suppresses the corresponding
decay widths much heavily, the decay widths are very small.
Considering the uncertainties of the masses, the decay widths of
the OZI allowed decay channels: $B_{s0}\to B\bar K$, $B'_{s1}\to
B^*\bar K$ are larger than the results of $B_{s1}$ and $B_{s2}$.
This is because that the $S$ doublet states ($0^+$, $1^+$) decay
through $S$ wave, they usually have broader decay widths, while
the $T$ doublet states ($1^+$, $2^+$) decay through $D$ wave, they
usually have narrower widths, though the decay channels are OZI
rule allowed ones.

The large discrepancies between the results of different models
may be caused by the small phase space of transition channels, and
the results are very sensitive to the masses of corresponding
mesons. We take into account the errors by varying all the input
parameters simultaneously within $\pm 5 \%$ of the central values.
One can see that in Table.~II, we get relatively large errors even
we change the parameters in such small regions. In Table.~III and
IV, the phase spaces of the decays of $B_{s1}$ and $B_{s2}$ are
too narrow, the decay widths depend heavily upon the masses of
initial mesons. If we change the masses of initial mesons, there
will be large errors, even one order larger than the original
value. So we change all the input parameters except the masses of
initial mesons which are fixed to the experimental data to
calculate the errors for $B_{s1}$ and $B_{s2}$.

In conclusion, through the improved B-S method, we predict the
masses of the orbitally excited states $B_{s0}$ and $B'_{s1}$,
calculate the strong coupling constants $G_{B_{s0}B_s\pi}$,
$G_{B_{s0}B\bar K}$, $G_{B'_{s1}B_s^*\pi}$, $F_{B'_{s1}B_s^*\pi}$,
$G_{B'_{s1}B^*\bar K}$, $F_{B'_{s1}B^*\bar K}$, $G_{B_{s1}B^{*}\bar
K}$, $F_{B_{s1}B^{*}\bar K}$, $G_{B_{s2}B\bar K}$,
$G_{B_{s2}B^{*}\bar K}$, and obtain the strong decay widths of
$B_{s0}\to B_s \pi$, $B_{s0}\to B\bar K$, $B'_{s1}\to B_s^* \pi$,
 $B'_{s1}\to B^*\bar K$, $B_{s1}\to B^*\bar K$, $B_{s2}\to B\bar K$ and $B_{s2}\to B^*\bar
K$, which are useful to find the unobserved states and to estimate
the full decay widths of these orbitally excited states.

\section*{ Acknowledgments}

This work was supported in part by the National Natural Science
Foundation of China (NSFC) under Grant No. 10875032, No.~11175051,
and supported in part by Projects of International Cooperation and
Exchanges NSFC under Grant No. 10911140267.

\end{document}